\documentclass[preprint,12pt, a4paper]{elsarticle}

\usepackage{amssymb}

\usepackage{lineno}

\usepackage{float}
\restylefloat{table}

\journal{SoftwareX}

\begin{document}

\begin{frontmatter}



\title{Optimized implementation of the conjugate gradient algorithm for FPGA-based  platforms \\ using the Dirac-Wilson operator as an example}


\author{G. Korcyl}

\address{Department of Information Technologies, Faculty of Physics, Astronomy and Applied Computer Science, Jagiellonian University, Krak\'ow, Poland}

\author{P. Korcyl}

\address{M. Smoluchowski Institute of Physics, Jagiellonian University, Krak\'ow, Poland}
\address{Institut f\"ur Theoretische Physik, Universit\"at Regensburg, Regensburg, Germany}

\begin{abstract}
It is now a noticeable trend in High Performance Computing that the systems are becoming more and more heterogeneous. Compute nodes with a host CPU are being equipped with accelerators, the latter being a GPU or FPGA cards or both. In many cases at the heart of scientific applications running on such systems are iterative linear solvers. In this work we present a software package which includes an FPGA implementation of the Conjugate Gradient algorithm using a particular problem of the Dirac-Wilson operator as encountered in numerical simulations of Quantum Chromodynamics. The software is written in OpenCL and C++ and is optimized for maximal performance. Our framework allows for a simple implementation of other linear operators, while keeping the data transport mechanisms unaltered. Hence, our software can serve as a backbone for many applications which are expected to gain a significant boost factor on FPGA accelerators. As such systems are expected to become more and more widespread, the need for highly performant FPGA implementations of the Conjugate Gradient algorithm and its variants will certainly increase and the porting investment can be greatly facilitated by the attached code.
\end{abstract}



\end{frontmatter}


%
%
%
%

\section*{Code Metadata}
\label{sec. version}

\begin{center}
\begin{table}[H]
\begin{tabular}{|l|p{6.5cm}|p{6.5cm}|}
\hline
\textbf{Nr.} & \textbf{Code metadata description} &  \\
\hline
C1 & Current code version & 1.0 \\
\hline
C2 & Permanent link to code/repository used for this code version & 
\footnotesize
\verb[https://bitbucket.org/[
\verb[fpgafais/hpcg_fpga[ 
\normalsize \\
\hline
C3 & Code Ocean compute capsule & None \\
\hline
C4 & Legal Code License   & GNU general public license\\
\hline
C5 & Code versioning system used & git \\
\hline
C6 & Software code languages, tools, and services used & C++, OpenCL \\
\hline
C7 & Compilation requirements, operating environments \& dependencies & Xilinx Vitis framework\\
\hline
C8 & Link to developer documentation/manual &  \verb[https://fpgafais.com/[
\verb[high-performance-computing/[ \\
\hline
C9 & Support email for questions & grzegorz.korcyl@uj.edu.pl \\
&& piotr.korcyl@uj.edu.pl\\
\hline
\end{tabular}
\caption{Code metadata}
\label{} 
\end{table}
\end{center}

\section{Motivation and Significance}
\label{sec. motivation}

High Performance Conjugate Gradient (HPCG) benchmark \cite{dongarra2016high} was included in the TOP500 ranking as a complementary way of measuring the performance of High Performance Computing (HPC) systems a couple of years ago. It was recognized that the High Performance LINPACK (HPL) \cite{hpl} benchmark based on dense matrix algebra does not represent the vast scope of applications whose performance relies more significantly on the interconnect between compute nodes. HP\-CG implements the Conjugate Gradient (CG) algorithm which is an iterative algorithm for solving large, sparse systems of linear equations. Such systems appear in many practical applications, for instance where partial differential equations are discretized. When the problem size reaches setups where parallelization over a large number of compute nodes is required, each node works is assigned its part of the solution. The boundary values have to be frequently exchanged between the neighbours as well as global communications are required to establish total error estimates. HPCG tries to replicate such communication patterns and provides benchmarks for their performance. 

Another application of iterative solvers arises in numerical simulations of Quantum Chromodynamics (QCD) \cite{gattringer}, where a large matrix representation of the Dirac-Wilson (DW) operator has to be frequently inverted. The sizes of the DW matrices often reach $10^9 \times 10^9$, implying that only the largest, parallel HPC systems can be realistically employed to handle this kind of problems. In all these cases it is evident that highly optimized implementations of the CG algorithm are extremely important for optimal usage of computer resources. 

The ever changing field of HPC has recently witnessed a new player come into the game. So far the most powerful HPC systems were constructed using two basic building blocks: many-core CPUs at the center of compute nodes and up to several highly-performant graphic cards (GPU) serving as accelerator cards. Currently, a third building block of a new architecture is appearing more and more frequently, namely FPGA-based accelerator cards, such as Xilinx Alveo cards \cite{alveo}. One of the  recent examples of such a highly heterogeneous HPC systems is the CYGNUS supercomputer at the Center for Computational Sciences University of Tsukuba in Japan \cite{cygnus}. 

FPGA processors operate on a different philosophy than CPUs or GPUs. They are not bounded by a fixed set of instructions and the entire program is translated into a specific configuration of logic gates. FPGA accelerators require a novel programming approach compared to the GPU cards and it is a very non-trivial task to port existing codes to such new platforms.

In this work we provide a highly optimized implementation of the CG algorithm for a single node FPGA accelerator. Although the implementation uses the Dirac-Wilson operator appearing in the context of QCD, it is generic and suitable for any other linear operator. This is because the Dirac-Wilson operator is defined on a 4D grid of points and its entries are complex matrices whose elements depend on the position in the grid. Hence, for each stencil not only the vector elements but also all the elements of the Dirac-Wilson matrix have to be loaded from the memory. One can consider this example as one of the most complex use cases of the CG algorithm. Therefore the presented implementation can be used as the backbone for any scientific application based on iterative solvers which needs to be ported to the new type of HPC systems. 

The particular problem solved by the presented code has not been considered in the literature at this level of sophistication as far as FPGA architectures are concerned. Single attempts were described in Ref.~\cite{4100953} and Ref.~\cite{janson}. Current implementation is based on our previous work presented in Ref.~\cite{Korcyl:2018pjc} and Ref.~\cite{Korcyl:2019uli}. We implemented a particular variant of CG algorithm which uses two data types as proposed in Ref.~\cite{strzodka}. For performance reasons we use cyclic buffers as suggested in Ref.~\cite{cyclic} in other context.

\section{Overview of Capabilities}
\label{sec. capabilities}

From the point of view of an HPC system, FPGA devices can, in most cases, be treated as ordinary accelerator cards. This means that a specific, highly compute demanding kernel must be singled out of the entire application and assigned to the FPGA for execution. Provided that input and output data are transferred efficiently a significant boost factor can be achieved due to the internal structure of FPGA processors, see for instance Ref.~\cite{Duarte:2018ite} for application in particle physics.

The most significant features of FPGA devices are natural parallelism and pipelined flow of computations. Natural parallelism refers to the fact that in a given programmable logic area of an FPGA processor many instances of the same computing kernel can be created. Once input data are provided to each of them, they will trivially execute the computations in parallel. Independently of that, the compute flow can usually be constructed in such a way as to overlap several computations on different input data within a single instance of the kernel. Such pipelined computations can often be initiated every cycle provided the input data is available and keep the logic resources busy thus maximizing overall performance.

We further improve the performance by overlapping data transfer with computations. To this aim we construct a set of mechanisms which allow to start the computations right after the very first complete batch of input data arrives into the memory of the FPGA accelerator. The results are returned to the host using a separate channel just after the computations of this particular data set are finished. This streaming data transfer removes a limit on the possible size of the local amount of data analyzed by the single compute node. The data transfer is hidden behind the computations and becomes irrelevant for large data sets.

For a long time FPGA devices were considered to be inadequate for floating point precision calculations. This has changed recently with the increasing amount of resources available on each device as well as with the introduction of dedicated blocks for single and double precision arithmetics fulfilling IEEE754 standard. We take advantage of the flexibility offered by the FPGA processors and implement the CG algorithm with an arbitrary fixed or floating point type of data following the version of the algorithm presented in Ref.~\cite{strzodka}. This allows one to exploit two data types simultaneously, a high and a low precision type. Most iterations are performed in low precision, whereas the high precision calculations are used to correct for a possible bias. The specific type of low and high precision types can be adjusted according to the problem at hand and required precision of the solution.

In our implementation we employ all the four advantages offered by the FPGA devices. Furthermore, we build a data transfer mechanism based on the concept of cyclic buffers as suggested in Ref.\cite{cyclic}. The idea is to keep in the local memory of the programmable logic input data for the neighbouring stencils, hence lowering the pressure on the memory bandwidth. Thus, for a consecutive stencil one needs to transfer only the data which have never been visited, while retrieving the remaining input data from the cyclic buffer from the local memory. All of the above improvements allow us to achieve a significant increase in performance as compared to CPU or GPU architectures. At the same time we keep the power requirements several times lower than those of GPU accelerators, which is one of the characteristic features of the FPGA technology.

\section{Software Description}
\label{sec. description}

The software presented in this work runs on a single compute node, equipped with a host and an FPGA-based accelerator. It has therefore a natural partition into two parts: a part which runs on the host processing unit, either a CPU or an ARM processor and the compute intensive kernel which is delegated to the programmable logic. One of the crucial pieces included in the package are the mechanisms responsible for efficient data transfer between the host and the programmable logic and back using DMA burst transfers. 

The host and the programmable logic are programmed using C++. Using the latest Vitis development framework from Xilinx \cite{vitis} one can keep the high-level abstraction layer throughout the entire project. Thus, complex data structures, as well as type templates can be used and truly facilitate the programming of the programmable logic. The data transfer and the invocation of the FPGA kernel are controlled by the host within the OpenCL framework. Alternative pieces of control code ensure that the same program can be compiled and executed exclusively on CPU for debugging and reference benchmarking purposes.

The program is controlled through an input script which provides algorithm parameters as well as paths to the input and output data. The size of the problem and several kernel parameters are set at compile time through a set of \verb[#define[ pragmas for performance reasons.

\subsection{Software Architecture}
\label{sec. architecture}

The starting point for the CG algorithm is the host. In each iteration it prepares data sets to be transferred to the accelerator, invokes the logic kernel for matrix-vector multiplication and gathers back the results. Overlapped data transfers (input and output data on separate channels) and pipelined computing kernel allow to process stencils in a streaming mode.


\subsubsection{Host Part}

The host is charged with the handling of the solution, residuum and additional vectors needed in the CG algorithm. Having access to the full residuum vector, the stopping criterion is also be evaluated in the host part of the code.

\subsubsection{Programmable Logic Part}

\begin{figure}
\begin{center}
\includegraphics[width=0.95\textwidth,angle=0]{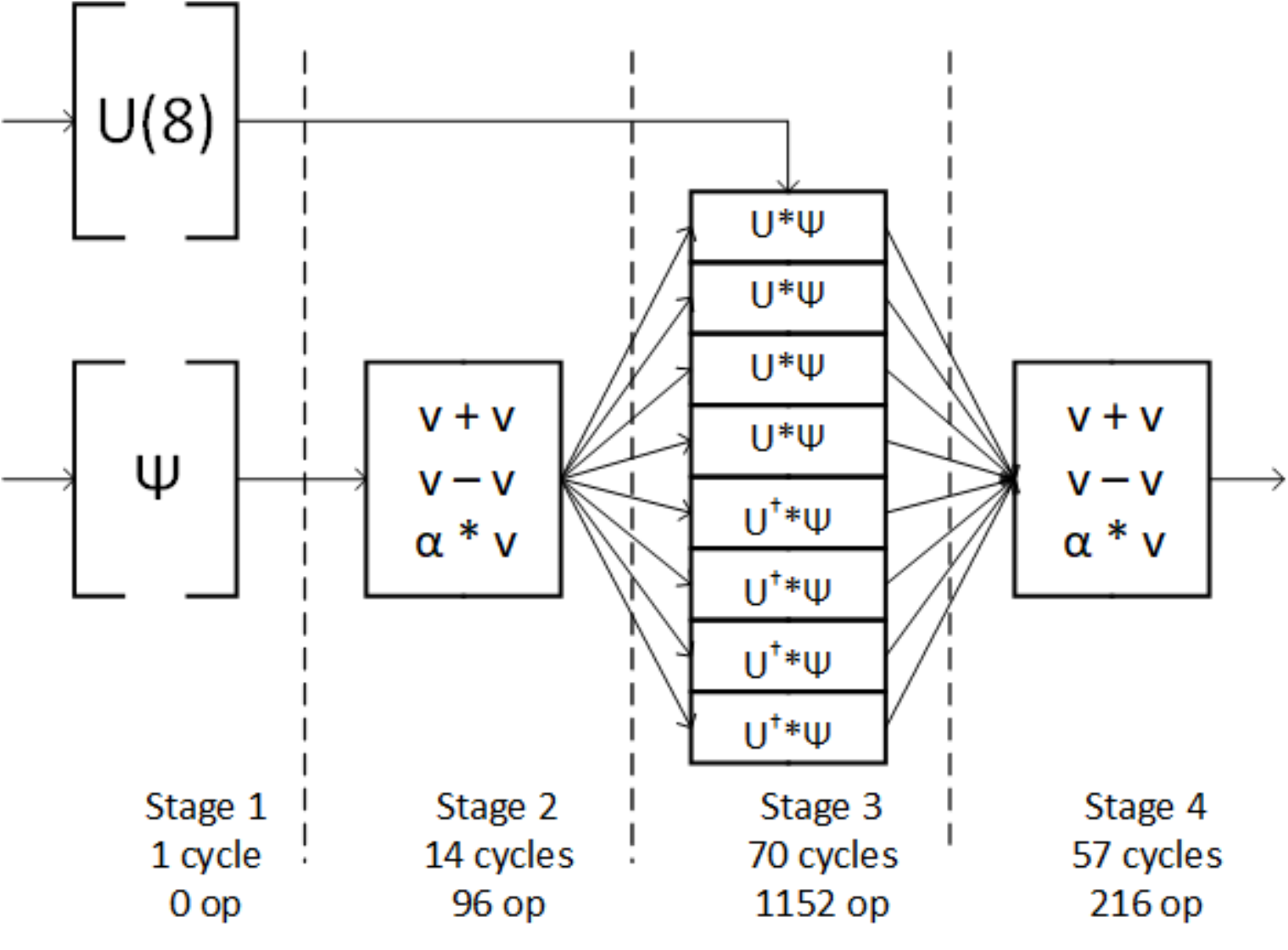}
\caption{Schematic view of the compute kernel implemented in programmable logic. The calculations can be divided into four stages. Each of them can be parallelized and pipelined. Due to the fix structure of calculations one can assign a fix number of clock cycles to each of the stages. The variables specific to the Dirac-Wilson operator are the 3x3 complex matrices $U$ and 12 component complex vectors $\psi$. In the first stage all required data is loaded in one clock cycle. All required linear algebra operators, such as linear combinations or matrix-vector multiplications are executed in parallel, as shown on the diagram in stages 2, 3 and 4. For different problems the input data for the stencil evaluation will be different, the amount of computation at each stage will be different, but the overall structure will remain. \label{fig. kernel}}
\end{center}
\end{figure}

The central part of the compute kernel executed in programmable logic is the function which evaluates the stencil. All possible computations are performed in parallel. Due to data interdependencies the final result is available after approximately 10 floating point operations, that is 142 clock cycles (given double precision). Moreover, the kernel has an initiation interval of one clock cycle, which means that the same infrastructure of logic gates can accept data for the evaluation of the next stencil in the consecutive clock cycle. Due to such architecture we maximize the use of compute resources and we are only limited by the memory bandwidth which limits the time needed to supply data for the next stencil. As already advertised, the latter is minimized by the use of cyclic buffers. We provide a schematic view of the function in Fig.~\ref{fig. kernel}, where we show all the calculations divided into four stages. The segmentation comes as a natural consequence of intermediate data dependence. 

\subsection{Software Functionalities}
\label{sec. functionalities}

The software package described in this work can be used to solve large, sparse sets of linear equations. It implements a variant of CG algorithm adapted to handle two, low and high precision, data types. The solution is found iteratively, with the main work executed at low precision, corrected occasionally by high precision iterations. The linear operator describing the set of equations is loaded from an input file which location is provided as an command line argument. The internal structure of the linear operator is hidden in the set of abstract C++ classes, which group the matrix and vector elements. It is hard-coded for performance reasons but can be easily adapted to any other problem at hand.


\section{Illustrative Examples}
\label{sec. examples}

\begin{figure}
\begin{center}
\includegraphics[width=0.95\textwidth,angle=0]{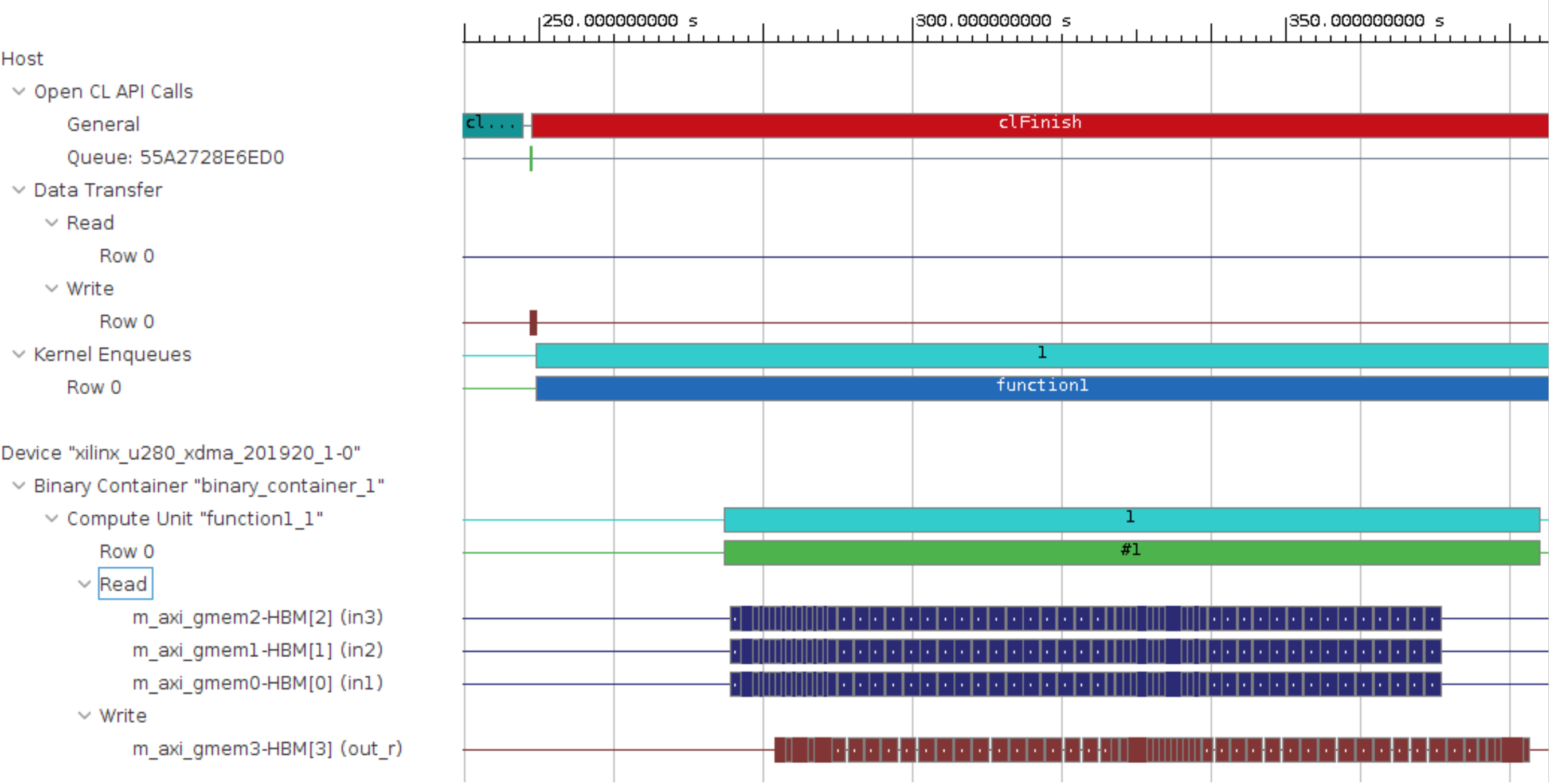}
\caption{Activity of various parts of the programmable logic during calculations as a function of elapsed time in microseconds of the simulation on the horizontal axis. Three violet blocks show the transfer of input data into the programmable logic through three memory channels. The brown block depicts the transfer of output data back to the host processor. Note the continuous nature of the four blocks and their overlap in time. \label{fig. trace}}
\end{center}
\end{figure}

As an example we show the program execution steps on a timeline. It allows to study the activity of various components in the system. In Fig.~\ref{fig. trace} we show the trace collected during the simulation process of a single kernel instance on Xilinx U280 Alveo accelerator card. The timeline shows simulation time and not execution time on real hardware. The red and light blue stripes show the complete processing time for the host to transfer the data from the DDR to FPGA memory, invoking the kernel and gathering back the results. In this example a blocking clFinish command is issued but a non-blocking mechanism can also be used and the host can proceed with other tasks during the kernel computations. The green stripe represents the kernel activity time, which includes streamlined reception of input data sets (3 HBM[2-0] channels, navy blue) and a single brown HBM3 channel which initiates data return after the kernel provides first results (kernel latency of 142 clock cycles). In our implementation the entire data transfer is hidden behind the computations and thus is invisible from the performance point of view. The uninterrupted usage of compute resources maximizes the performance.

\section{Performance}
\label{sec. performance}

We have benchmarked out implementation using the new Xilinx U280 Alveo card. The device has enough resources and data channels to fit in three fully-parallel instances of the kernel. In order to fully profit from the device architecture, the kernel was configured for \verb[float[ data type and initiation interval of 2 clock cycles running at 300 MHz. In such configuration achieved sustained performance levels at 607 GFLOPs which overpasses 64-cores Intel Xeon Phi processors \cite{Durr:2018ayq} and reaches Nvidia V100 GPU \cite{gpu} running similar algorithm.

\section{Impact}
\label{sec. impact}


The software presented in this work 
provides an example of the optimized implementation of the CG algorithm for systems equipped with FPGA-based accelerator cards. The construction of the compute kernel and of the data transfer mechanism ensures the maximisation of the usage of compute blocks and the minimization of the required data to be transferred, thus providing an optimal implementation which maximizes performance. Regarding the general context, the code showcases multiple techniques for implementing computing kernels on FPGA logic using OpenCL and C++ in order to profit from that technology advantages. These techniques include: various computations optimizations for reaching iteration interval of one clock cycle, efficient memory management and data placement, streamlined design and custom data types. As systems with FPGA-based accelerator cards are likely to become more and more widespread, our software can greatly facilitate the porting of the existing applications to new computing platforms. Moreover, our implementation can be also used as the stepping stone for the construction of multigrid algorithms which in many cases represent the state-of-the-art iterative solvers. Last but not least, any new FPGA-equipped system may require an adapted implementation of the HPCG benchmark in order to fully assess its performance. The presented code can be readily adapted to that purpose by tailoring the implemented host-FPGA accelerator card connection to the one actually realized in the system being benchmarked. 
%
%
Our package can serve as the starting point for the development of HPCG variants optimized for the HPC systems featuring FPGA accelerators, which offer a number of features that makes them superior versus GPU such as: configurable internal architecture, low power consumption, dynamic reconfiguration and low-level networking with short and deterministic latency.






\section{Conclusions}
\label{sec. conclusions}

In this work we presented a software package which implements the CG algorithm for FPGA-based accelerator cards. It provides an example of highly optimized code using an an example the Dirac-Wilson operator encountered in Monte Carlo simulations of elementary particles. The code can be easily adapted to handle other linear operators, thus providing a backbone for all applications which seek a performance boost after porting to FPGA-based architectures.

%


\section*{Acknowledgements}
\label{sec. acknowledgements}

This  work  was  in  part  supported  by  Deutsche  Forschungsgemeinschaft  under  Grant  No.SFB/TRR 55 and by the polish NCN grant No. UMO-2016/21/B/ ST2/01492, by the Foundation for Polish Science grant no. TEAM/2017-4/39 and by the Polish Ministry for Science and Higher Education  grant  no.  7150/E-338/M/2018.  The  project  could  be  realized  thanks  to  the  support from Xilinx University Program and their donations. P.K. acknowledges support from the NAWA Bekker  fellowship  and  thanks  Universita  degli  Studi  di  Roma  Tor  Vergata  for  hospitality  during which parts of this work were performed.



\bibliographystyle{elsarticle-num} 
\bibliography{ref}






\end{document}